\documentclass{vgtc}                          

\graphicspath{{figures/}{pictures/}{images/}{./}} 

\usepackage{times}                     

\usepackage{tabu}                      
\usepackage{booktabs}                  
\usepackage{lipsum}                    
\usepackage{mwe}                       

\usepackage{mathptmx}                  

\onlineid{1333} 

\vgtccategory{Research}

\vgtcinsertpkg

\usepackage{comment}
\usepackage{textcomp}
\usepackage{gensymb}

\title{SABER: Spatial Attention, Brain, Extended Reality\thanks{This is a preprint version of this article. The final version of this paper can be found in the Proceedings of IEEE VR 2026. For citation, please refer to the published version. This work was initially made available on the author's personal website [yujnkm.com] on February 2024, and was subsequently uploaded to arXiv for broader accessibility.}}

\author{Tom Bullock\textsuperscript{1,2,3}
\and Emily Machniak\textsuperscript{1,2,4}
\and You-Jin Kim\textsuperscript{5}
\and Radha Kumaran\textsuperscript{6}
\and Justin Kasowski\textsuperscript{1}
\and Apurv Varshney\textsuperscript{6} 
\and Julia Ram\textsuperscript{1}
\and Melissa M. Hernandez\textsuperscript{1}
\and Stina Johansson\textsuperscript{1}
\and Neil M. Dundon\textsuperscript{1,3}
\and Tobias Höllerer\textsuperscript{2,6}
\and Barry Giesbrecht\textsuperscript{1,2}}

\affiliation{
\vspace*{-1.5ex} 
\textsuperscript{1} \scriptsize{Department of Psychological and Brain Sciences, University of California, Santa Barbara, Santa Barbara, CA, United States.}\\
\textsuperscript{2} \scriptsize{Institute for Collaborative Biotechnologies, University of California, Santa Barbara, Santa Barbara, CA, United States.}\\
\textsuperscript{3} \scriptsize{Department of Social and Psychological Sciences, University of Huddersfield, Huddersfield, United Kingdom}\\
\textsuperscript{4} \scriptsize{Department of Human Systems Engineering, Arizona State University, Mesa, AZ, United States.}\\
\textsuperscript{5} \scriptsize{School of Computing, University of Nebraska--Lincoln, Nebraska, United States.}\\
\textsuperscript{6} \scriptsize{Department of Computer Science, University of California, Santa Barbara, Santa Barbara, CA, United States.}\\

\vspace*{-3ex} 
}

\teaser{
  \centering
  \includegraphics[width= 0.90\linewidth]{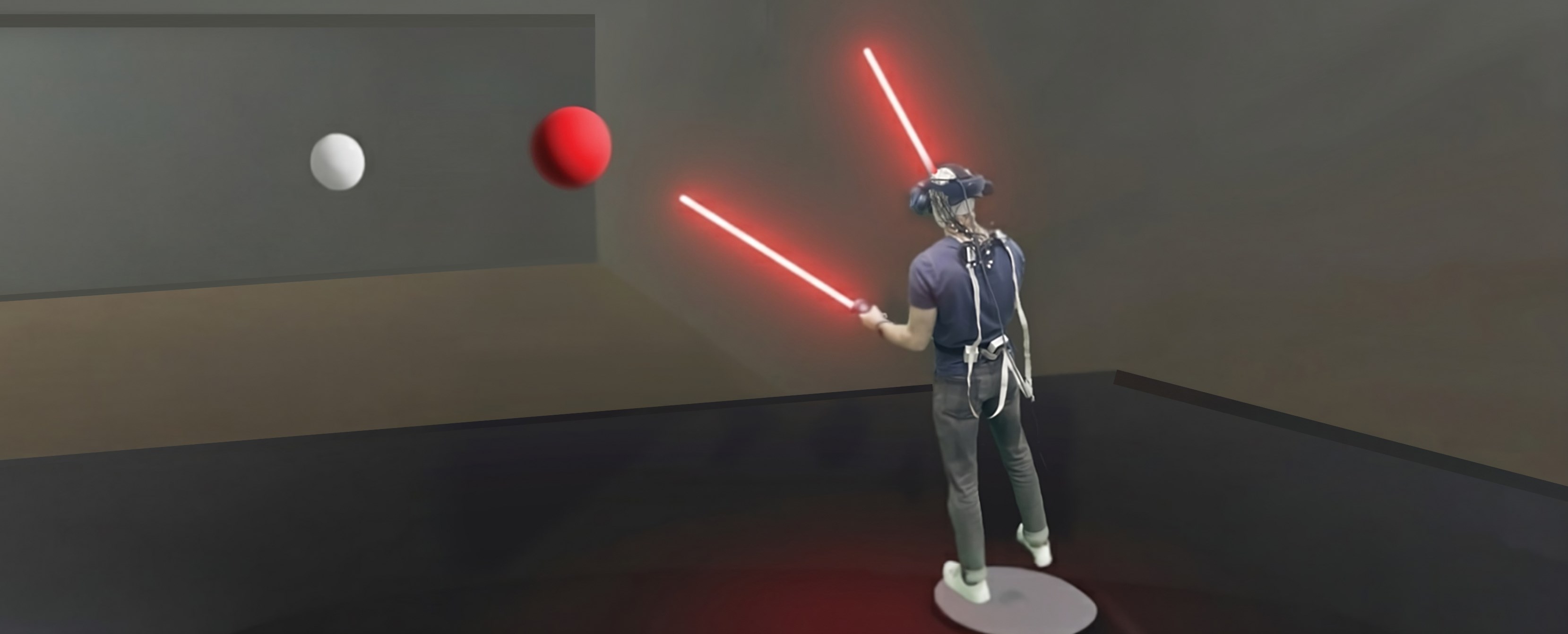}
  \caption{A participant is shown superimposed into a virtual environment, standing on a platform wielding a pair of virtual sabers anchored to HTC Vive Controllers. They are required to strike incoming spheres using either saber while their brain activity is recorded at the scalp using electroencephalography (EEG).}
  \label{fig:Fig01_Teaser} 
}

\abstract{
Tracking moving objects is a critical skill for many everyday tasks, such as crossing a busy street, driving a car or catching a ball. Attention is a key cognitive function that supports object tracking; however, our understanding of the brain mechanisms that support attention is almost exclusively based on evidence from tasks that present stable objects at fixed locations. Accounts of multiple object tracking are also limited because they are largely based on behavioral data alone and involve tracking objects in a 2D plane. Consequently, the neural mechanisms that enable moment-by-moment tracking of goal-relevant objects remain poorly understood. To address this knowledge gap, we developed SABER (Spatial Attention, Brain, Extended Reality), a new framework for studying the behavioral and neural dynamics of attention to objects moving in 3D. Participants (\textit{n}=32) completed variants of a task inspired by the popular virtual reality (VR) game \textit{Beat Saber}, where they used virtual sabers to strike stationary and moving color-defined target spheres while we recorded electroencephalography (EEG). We first established that standard univariate EEG metrics which are typically used to study spatial attention to static objects presented on 2D screens, can generalize effectively to an immersive VR context involving both static and dynamic 3D stimuli. We then used a computational modeling approach to reconstruct moment-by-moment attention to the locations of stationary and moving objects from oscillatory brain activity, demonstrating the feasibility of precisely tracking attention in a 3D space. These results validate SABER, and provide a foundation for future research that is critical not only for understanding how attention works in the physical world, but is also directly relevant to the development of better VR applications. The insights gained here can potentially inform the design of more intuitive interfaces, effective training simulations, and immersive experiences optimized for the human attention system.
}

\begin{document}

\firstsection{INTRODUCTION}
\maketitle 

Attending to moving objects, continuously updating their positions, and predicting their trajectories are critical for goal-directed actions in everyday life. In the physical world, simply walking down a busy street requires us to navigate amongst multiple moving objects, such as other pedestrians, cyclists and vehicles. This challenge extends to the digital world. When playing a VR first-person shooter, one must track and target an enemy player while simultaneously monitoring other players on the battlefield. Similarly, in a rhythm game such as \textit{Beat Saber}, players must track multiple incoming targets and anticipate their positions to prepare and execute precise actions. The human capacity to construct stable mental representations that support the tracking of one, or even multiple, dynamic objects is well-documented ~\cite{pylyshyn_tracking_1988,holcombe_attending_2023,cavanagh_tracking_2005}, while impaired object tracking is often associated with various brain disorders and injuries ~\cite{alnawmasi_deficits_2022,harenberg_effectiveness_2021,kelemen_how_2007}. Given the fundamental importance of object tracking in both our physical and digital worlds, it is critical that we understand the brain mechanisms that enable this function.

Evidence from cognitive psychology and neuroscience suggests that our ability to track objects is heavily dependent on visual attention and working memory (WM)~\cite{belledonne_adaptive_2025, drew_neural_2008, holcombe_attending_2023}. These processes support selective sampling, or prioritizing, of behaviorally relevant sensory evidence from the environment, integrating this evidence with expectations about the object, maintaining the representation over time, and then potentially executing a behavioral action ~\cite{drew_neural_2008}. Unfortunately, current accounts of visual attention and working memory are almost exclusively based on evidence from tasks using stable objects at fixed locations ~\cite{Corbetta2002, fiebelkorn_functional_2020, luck_progress_2021, shomstein_attention_2023, wolfe_guided_2021,awh_working_2025} and there is limited evidence about the moment-by-moment neural dynamics that support the tracking of dynamic objects. Neural explanations from multiple object tracking (MOT) studies are also limited because these studies are largely based on behavioral data alone and involve tracking objects in a 2D plane ~\cite{cavanagh_tracking_2005, holcombe_attending_2023, pylyshyn_tracking_1988,dedrick_perception_2009}. Thus, despite the fundamental importance of tracking dynamic objects in the environment, and the disruption of this capacity by disease and injury, our understanding of the perceptual, cognitive and action-related mechanisms underlying these processes is limited. 

To address this knowledge gap, we present a new approach, inspired by \textit{Beat Saber}, which combines VR and EEG to study the neural mechanisms of spatial attention to dynamic objects in an immersive 3D environment. We refer to this approach as SABER (Spatial Attention, Brain, Extended Reality). We validate this approach with results from an experiment where a sample of healthy young adults (\textit{n}=32) used virtual sabers to strike color-defined target spheres while we recorded electrical brain activity at 64 scalp locations using EEG. Targets appeared in one of two ways: they either emerged directly in front of the participants or moved toward them from a distance. These targets were presented either alone or surrounded by an array of distractors (different colored spheres that participants were instructed to ignore). 

We first demonstrate that standard univariate EEG methods---typically used to study spatial shifts in attention to static objects presented on 2D screens---can also be successfully applied to analyze brain responses to both static and dynamic objects in immersive 3D VR. We then apply a multivariate computational method to show that both static and moving object locations and trajectories can be reconstructed from patterns of alpha brain activity recorded at the scalp. Together, these results validate SABER, opening up exciting future opportunities to test hypotheses about how the brain supports attention to dynamic objects. The potential insights that can be gained from this new approach to studying attention are critical for two key reasons. First, they provide insights into how attention operates in dynamic scenarios in the physical world. Second, they will aid the development of better virtual and mixed-reality applications, informing the design of more intuitive interfaces, effective training simulations, and immersive experiences that are optimized for the human attention system. Ultimately, the insights gained from this work have broad interdisciplinary relevance, with implications for cognitive neuroscience and psychology, human factors, and the development of extended reality (XR) applications and brain-computer interfaces (BCI).

\section{RELATED WORK}

\subsection{Multiple Object Tracking} 
\label{subsec: MOT}
In both the physical and virtual worlds, achieving our goals and avoiding potential dangers often requires that we attend to and track moving objects. This ability relies heavily on spatial attention and WM ~\cite{belledonne_adaptive_2025,drew_neural_2008,holcombe_attending_2023}, and the majority of the evidence implicating visual attention and visual working memory comes from studies using the MOT task ~\cite{belledonne_adaptive_2025, drew_neural_2008, holcombe_attending_2023}. In this simple computer task, participants are typically shown multiple objects and cued to track one or more objects as they move on a computer screen. The MOT literature is large and has provided important insights into attention and WM. Spatial attention supports the initial selection of the targets to be tracked, while WM supports the sustained representation of behaviorally relevant objects ~\cite{allen_multiple-target_2006,drew_neural_2008,meyerhoff_studying_2017,zhang_objects_2010}. However, accounts of object tracking based on MOT are limited primarily because they largely involve participants tracking objects on a 2D plane (but see ~\cite{thomas_self-motion_2010}) and are typically based on behavioral data alone, with relatively few studies also recording brain data (~\cite{drew_neural_2008}). Indeed, while the MOT task provides an excellent methodology for studying the behavioral dynamics of object tracking, the task by its very nature is not well-suited for studying the neural dynamics of tracking a specific object. 

Contemporary theories of attention and WM provide limited insight on the tracking of dynamic objects because they are primarily based on static objects presented at fixed locations ~\cite{Corbetta2002,fiebelkorn_functional_2020,luck_progress_2021,shomstein_attention_2023,wolfe_guided_2021}. Consequently, there is a significant gap in our understanding of how the brain handles dynamic stimuli. Indeed, simplifying attention to the perception of still, 2D images overlooks that attention must operate dynamically over time to prioritize relevant information and use it when most needed ~\cite{denison_visual_2024}.

SABER is specifically designed with the goal of analyzing both the behavioral and neural dynamics of moment-by-moment object tracking. This will allow for the moment-by-moment modeling of attention to a specific goal-relevant 3D target object that is either tracked within the environment in isolation or amongst multiple other goal-irrelevant objects. The hope is that this study and future work based around this approach will provide insight into how known mechanisms from static object processing extend to contexts involving dynamic interactions with 3D moving objects. This work will inform contemporary models of spatial attention and WM. 



\subsection{EEG Approaches to Studying Spatial Attention} %
EEG is a method for noninvasive measurement of brain activity from electrodes placed on the scalp ~\cite{berger_uber_1929}. The measured signal captures post-synaptic electrical activity from large populations of neurons that are synchronously activated ~\cite{10.1093/acprof:oso/9780195050387.001.0001}. The acquired signal has a high temporal resolution and is a direct measure of neural activity. In cognitive neuroscience, there are two main features of the EEG signal that have been used to inform the fundamental understanding of sensory, perceptual, and cognitive phenomena: time-varying amplitude and oscillatory power.


\emph{Time-varying amplitude}. To extract the voltage associated with specific task events, EEG studies have used the Event-Related Potential (ERP) technique~\cite{luck2014introduction}. This technique involves averaging EEG activity that has been time-locked to an event, such as an shape appearing on a screen. By averaging across many repetitions of the event (i.e., trials), random background activity in the EEG signal is canceled out, and what remains is a time-varying waveform that represents the neural response to the specific event. Additional analytical steps can then be taken to control for stimulus differences and provide more anatomical specificity by leveraging the known systematic mapping of the visual field in cortical visual areas~\cite{wandell_visual_2007}. In ERP studies of visual attention, one common step leverages the contralateral organization of visual cortex ~\cite{luck2014introduction} by comparing brain activity over electrodes contralateral (i.e., opposite side) to the location of an attended stimulus versus activity in electrodes ipsilateral (i.e., same side).  The difference between these two signals is referred to as \textit{lateralization} and has been leveraged to gain insight into the neural mechanisms that underpin spatial attention. One particularly useful ERP component that is based on this approach is the N2pc, which is a consistently larger negative deflection of the waveform between around 200-350 ms at posterior electrode sites contralateral to an attended stimulus and is thought to reflect the enhancement of the target signal from distracting information ~\cite{eimer_n2pc_1996,luck_electrophysiological_1994}. While the ERP technique provides valuable insight into the mechanisms of spatial attention, a key limitation is that it is typically only be applied to stimuli that have an abrupt onset. This makes it unsuitable for studying the moment-by-moment fluctuations of attention over time, as it cannot be used with gradual or continuous stimuli.

\emph{Oscillatory power}. The EEG signal can also be converted into the frequency domain to quantify oscillatory activity and this approach is also well-suited to capturing the neural dynamics of spatial attention. In this approach, the EEG signal is converted into the frequency domain using various transformation methods, including Fourier, Hilbert, or Morelet wavelet transforms ~\cite{cohen_analyzing_2014}. It is well-established that activity in the alpha frequency band ($\sim$8-13 Hz) tracks both the locus and timing of covert spatial attention ~\cite{thut_alpha-band_2006, sauseng_shift_2005}. Specifically, alpha builds up at posterior sites ipsilateral to the attended location, which has lead to the view that it reflects gating/suppression of irrelevant visual information ~\cite{foxe_role_2011}, although it could also reflect signal enhancement ~\cite{foster_role_2019}. Examining lateralized alpha responses provides insight into moment-by-moment fluctuations in attention over time, which means it can be applied to gradual or continuous stimuli. However, lateralization is still a relatively crude metric in the sense that it limits our understanding of spatial attention to "attend left vs. attend right" and does not provide insight into the quality of the representation. 

Amplitude and oscillatory features have traditionally been analyzed using univariate statistical techniques, but more recently, researchers have been using multivariate analytic techniques to investigate patterns of brain activity. One technique used in cognitive neuroscience is the Inverted Encoding Model (IEM). Originally developed for fMRI to recover feature-selective responses in visual cortex ~\cite{brouwer_decoding_2009}, this method assumes that the aggregate signal measured at the scalp (EEG) reflects the summed activity of underlying neural populations, each governed by specific tuning properties (e.g., spatial receptive fields). Specifically, the goal of IEM is to determine the extent to which the data conform to an \emph{a priori} formulation of what the response to the stimulus conditions should look like. This methods allows researchers to visualize the "quality" or "fidelity" of the population-level representation of the stimulus directly. In the context of attention, IEMs have been used to demonstrate that topographic patterns of alpha-band oscillations can be used to reconstruct spatially selective neural responses to both attended object orientations ~\cite{garcia_near-real-time_2013, bullock_acute_2017} and spatial locations ~\cite{samaha_decoding_2016}, as well as their encoding and maintenance ~\cite{foster_topography_2016,maclean_dual_2019,garrett_tracking_2021,bullock_eye_2023}. Crucially for our study, IEMs applied to scalp EEG have proven robust to noise artifacts that are common in physically active tasks, including motion artifacts and sweating ~\cite{bullock_acute_2017,garrett_tracking_2021}.

In this paper, we apply both univariate (ERP, lateralization) and multivariate (IEM) techniques—typically used for stationary objects on 2D screens—to EEG data collected during our task. Our goal is to determine the extent to which these methods can be adapted to study spatial attention to both static and moving objects in an immersive 3D VR environment.

\subsection{Virtual Reality and Mobile Brain Imaging}

A key goal of cognitive neuroscience is to understand the brain mechanisms that support natural human behavior. By building models of how we acquire, process, store and apply information, these findings can help solve real-world problems, such as designing better interfaces for technology, developing therapeutic interventions for cognitive disorders and improving educational techniques. Several decades of laboratory research have provided valuable insights into the human brain and behavior. However, traditional laboratory experiments are often a poor reflection of reality. These experiments typically rely on stationary 2D stimuli presented to participants on a flat screen, requiring simple responses that are disconnected from natural responses (e.g., pressing buttons on a keyboard, clicking a mouse). This approach delivers high internal validity, but limited ecological validity, meaning that the neural mechanisms underlying natural behavior are unclear ~\cite{thurley_naturalistic_2022}. Furthermore, traditional brain imaging studies have been largely conducted on sedentary participants who are required to remain as still as possible to minimize movement artifacts. As a result, current cognitive models largely fail to account for the impact of movement on brain activity, despite increasing evidence that physical activity can have profound effects on multiple aspects of brain and behavior~\cite{giesbrecht_physically_2025,garrett_systematic_2024}. Indeed, recent work has shown that even the simple act of standing versus sitting while completing a task can impact task-specific brain activity~\cite{fakorede_neural_2025,zinchenko_withstand_2025}. Together, these limitations in traditional laboratory based tasks highlight a significant knowledge gap in cognitive neuroscience regarding the neural basis of natural human behavior in dynamic, real-world conditions.

A combined VR and mobile brain imaging approach offers a promising solution to this problem, allowing for the study of more naturalistic behavior ~\cite{stangl_mobile_2023,thurley_naturalistic_2022}. VR offers advantages over traditional lab methods because it offers greater ecological validity while maintaining high levels of experimental control. VR uniquely combines environmental immersion with experimental rigor: it allows participants to interact with complex, dynamic 3D stimuli within a standardized framework where events can be precisely replicated across trials. Concurrently, advances in mobile EEG technology allow for non-invasive, wireless recording of brain activity from active humans, with numerous studies successfully demonstrating that movement and environmental noise artifacts can be minimized to achieve a good signal-to-noise ratio (e.g., ~\cite{wang_human_2024, delaux_mobile_2021}). Although studies directly comparing VR with traditional 2D displays are sparse, emerging evidence suggests that VR can elicit even more potentiated brain responses ~\cite{schubring_virtual_2020,schone_reality_2023}, indicating that a combined VR-EEG approach is a promising method for studying naturalistic human brain and behavior.

Our new approach, SABER, combines VR and EEG to allow for multimodal data sampling from freely moving participants in an immersive environment. This provides a platform to study the neural mechanisms that underpin tracking and engaging with dynamic 3D objects, in a reproducible, yet more naturalistic manner.

\begin{figure}[t] 
  \centering \includegraphics[width = \columnwidth]{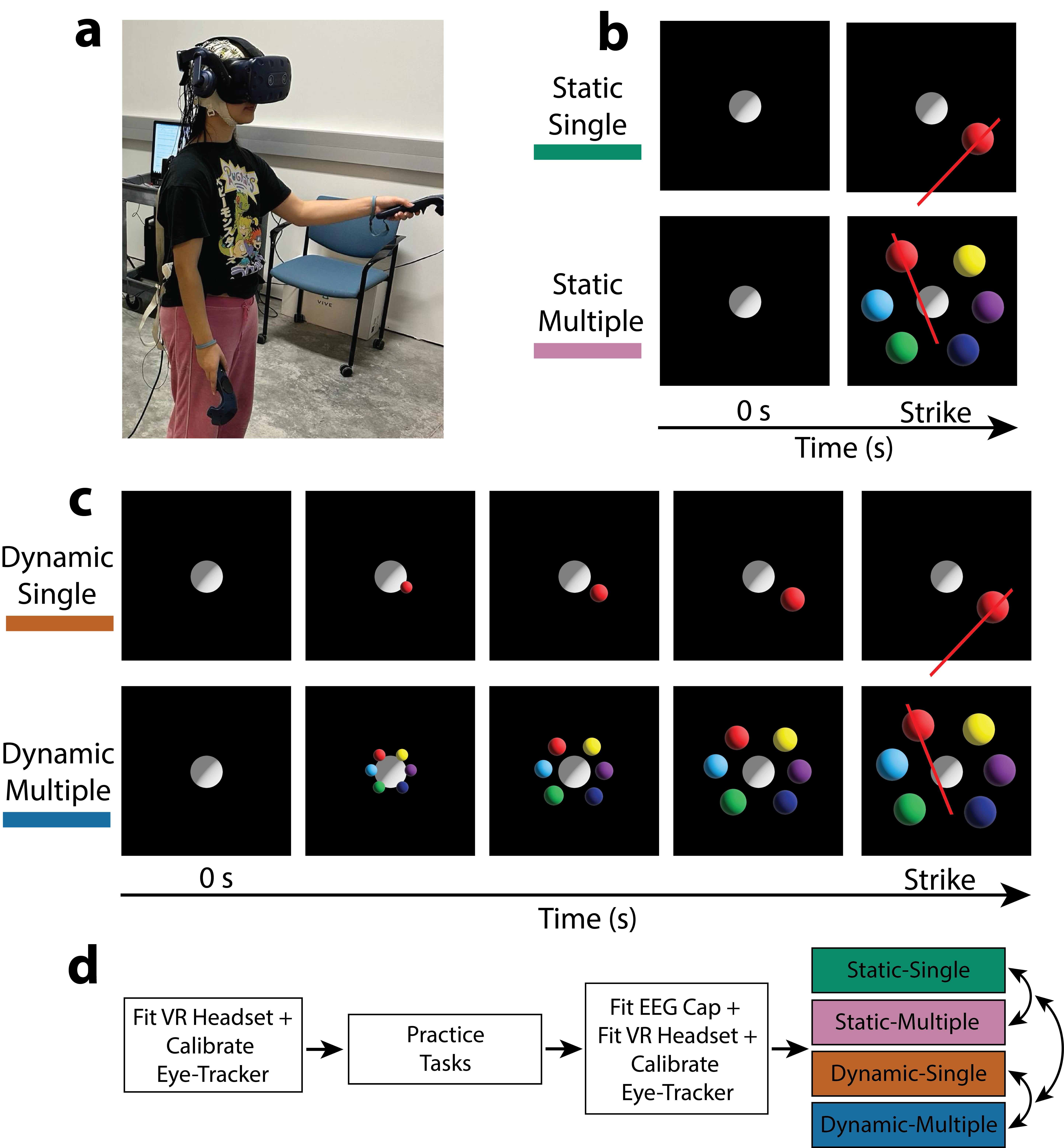}
  \caption{VR-EEG setup, cognitive tasks and protocol. (a) A participant engaged in the task while wearing an EEG cap and VR headset. (b) Schematic examples of trials in the Static- Single and Multiple conditions. A red target sphere appeared in front of the participant either in isolation or flanked by non-targets and the participant used either saber to strike the target as quickly as possible. (c) Schematic examples of trials in the Dynamic- Single and Multiple conditions.  On each trial a target sphere appeared at a distance, either in isolation or flanked by non-targets, and moved towards the participant. They used either saber to intercept the target when it came within reach. (d) Experimental workflow. Curved arrows depict the between-participants counterbalancing of the four main conditions.} 
  \vspace{-2ex} 
  \label{fig:Fig02_Methods}
\end{figure}

\section{METHODS}

\subsection{Participants} 
Thirty-three healthy adults (24 female, 32 right-handed, $M_{age}$ = 22.12, $SD_{age}$ = 3.10) were recruited. Participants reported normal or corrected-to-normal vision. Informed consent was provided at the beginning of each session. All procedures were approved by the UC Santa Barbara Ethics Committee and the U.S. Army Human Research Protection Office. Participants received \$20/hr for time spent in the lab ($\sim$2.5 hrs total per participant). 

The sample size was based on a power analysis which revealed that with a significance level of $\alpha$ = .05 and power of .80, we needed a minimum of 34 participants to detect a medium effect and 15 participants to detect a large effect. We aimed to recruit 32 participants as this allowed us to counterbalance the order of the four experimental conditions (see next section), but ended up recruiting an extra participant due to a technical error during one recording resulting in no trigger codes being sent, hence 33 participants total. With the participant with the technical error excluded, the final sample was 32 participants. We further reasoned that this sample size was sufficient for validating our approach, as it meets or exceeds those of foundational studies using these methods. For instance, seminal work on the N2pc ~\cite{eimer_n2pc_1996}, alpha lateralization ~\cite{sauseng_shift_2005}, and Inverted Encoding Models ~\cite{samaha_decoding_2016} relied on sample sizes of 10–11, 22, and 8 subjects, respectively. 


\subsection{Task and Procedure} 
The experimenter welcomed the participant, providing a detailed explanation of the study and obtaining their informed consent. The participant was then equipped with the VR headset (HTC Vive Eye Pro, HTC Vive, Taoyuan City, Taiwan) and controllers, and guided through single practice blocks of the "Dynamic-Multiple" and "Dynamic-Single" conditions (see below) of the task. All participants successfully completed the practice blocks and proceeded to the full experiment.

The participant then removed the VR headset, sat in a comfortable chair, and was fitted with an EEG cap consisting of 64 scalp electrodes arranged according to the 10-20 system (ActiChamp, Brain Products, Berlin, Germany). Two electrodes were placed directly on the mastoids. Electrode FCz was used as an online reference. A wireless EEG transmitter (MOVE, Brain Products) was then attached. The participant then stood up and the experimenter positioned the VR headset over the EEG cap (see Figure \ref{fig:Fig02_Methods}A) and handed them a Vive controller for each hand. This VR-EEG setup is consistent with previous studies ~\cite{delaux_mobile_2021,liang_dissociation_2018}.

The participant then completed the main experimental task while we continuously sampled EEG at 1000 Hz. The task was controlled using custom scripts for Unity (Unity Technologies, San Francisco, USA) that were written with functions from the Simple XR toolbox  (https://github.com/simpleOmnia/sXR) ~\cite{kasowski_simplexr_2023}. A Dell XPS Desktop 8960 PC (NVIDIA RTX4090 GPU, Intel i9 Processor and 64 GB Ram) was used to run the VR setup. Event timing codes were sent from the Unity PC to the EEG amplifier using custom code for Unity. Eye-tracking data from the Vive headset were recorded throughout the experiment at 90 Hz, as were the positions of all relevant objects in the environment. Prior to completing the first condition, the participant ran through the built-in Vive eye-tracker calibration routine, where the physical inter-pupillary distance (IPD) was manually adjusted and headset position optimized. All participants successfully completed the calibration routine, confirming adequate visual clarity and alignment according to the Vive software (mean IPD = 59.9 mm, SD IPD = 3.1 mm). 

The participant was first positioned on a platform in a dark, sparse virtual room (Figure ~\ref{fig:Fig01_Teaser}). They were handed a pair of Vive controllers, which transformed into virtual sabers measuring 1.5 m in length, the end of each saber anchored to a controller. Both sabers were the same color as the targets (counterbalanced red/blue between participants). They were instructed to position themselves over a white circle on the platform and maintain fixation on a fixation sphere (2 m circumference) positioned at a distance of 30 m and a height of 1 m. They then completed blocks of the four task conditions (order counterbalanced between participants). Each block was initiated with a trigger press, and was comprised of a continuous stream of stimuli. 

In the ~\textit{Static-Single} condition, on each trial a lone target sphere (0.4 m circumference) abruptly appeared in one of six possible pseudo-random locations (centered 1 m from the fixation sphere) and they were instructed to strike the sphere as quickly as possible using either saber (Figure \ref{fig:Fig02_Methods}B). The target color was either red or blue, counterbalanced between participants. If the target was struck by a saber within the response window (1 s), it disappeared, and the participant received haptic feedback (a short buzz on the striking controller) to signal their successful target strike. The next target then appeared in another pseudo-random location, and the sequence continued. In the ~\textit{Dynamic-Single} condition, the target sphere emerged from a distant fixation point and moved towards one of the six location bins, and participants were instructed to strike it as it traveled past them (Figure \ref{fig:Fig02_Methods}C). At the point of saber contact, the moving target sphere was approximately the same distance from fixation as in the Static condition (1 m). Successful saber strikes triggered a short confirmatory buzz on the striking controller. The ~\textit{Static-Multiple} and ~\textit{Dynamic-Multiple} conditions were identical to the Static-Single and Dynamic-Single conditions in terms of trial structure and timing, but on each trial five distractor spheres appeared at the same time as the target sphere,  one in each of the remaining location bins. The distractor colors were green, magenta, cyan, yellow and either blue if the assigned target color was red, or red if the assigned target color was blue. 

The participant was instructed to maintain gaze on the fixation sphere and minimize eye and head movements throughout each trial block. Fixation breaks caused the sphere to turn from white to grey, which was intended to prompt the participant to re-fixate on the sphere. Fixation breaks were defined as gaze deviations exceeding 5 degrees of visual angle from the center of the fixation sphere. In all conditions the target destination was one of six different stimulus locations, which were determined by pseudorandomly selecting one of six locations [30\degree, 90\degree, 150\degree, 210\degree, 270\degree, 330\degree] and then adding a random amount of jitter [+/- 10\degree]. Target location sequences were pseudo-randomized to ensure equal sampling across bins while preventing repetition of the same bin on consecutive trials. A unique sequence was generated for each participant and condition. Each condition consisted of 6 blocks, and each block contained 102 trials. Condition order was counterbalanced across four possible permutations to control for sequence effects (SS = Static-Single, SM = Static-Multiple, DS = Dynamic-Single, DM = Dynamic-Multiple): [SS,SM,DS,DM; SM,SS,DM,DS; DS,DM,SS,SM; DM,DS,SM,SS]. Participants were distributed across these four orders. They were given the opportunity to take breaks between blocks during each condition, and also longer breaks between conditions, including temporarily removing the VR headset if required. In the static conditions, the 1 s response window was determined via pilot testing. All four conditions took $\sim$1 hr total to complete. Participants tolerated the experimental protocol well, and no sessions were terminated due to simulator sickness or adverse reactions.

\subsection{EEG Data Preprocessing}
\label{subsec:EEG_Data_Preprocessing}
EEG data were processed offline with custom scripts that use functions from the EEGLAB Toolbox ~\cite{delorme_eeglab:_2004} and MATLAB (version 2023a, The MathWorks, Inc., Natick, MA, US). First, EEG event markers were shifted to compensate for the latency between the event codes sent from Unity to the amplifier to mark the onset of each stimulus on each trial, and the actual rendering of each stimulus in the VR headset. This correction was based on independent validation using a photodiode, which revealed a mean lag of 25.56 ms ($SD = 2.02$ ms) across 102 trials. The EEG data were downsampled to 250 Hz to reduce processing and memory demands. The data were then re-referenced to the average mastoid signal, and low- and high- pass filters were applied at 30 Hz and 0.1 Hz, respectively. The Automatic Artifact Removal (AAR) toolbox ~\cite{gomez-herrero_automatic_2006} was then applied to remove ocular artifacts. Specifically, ocular artifacts were corrected using the AAR toolbox’s conventional recursive least squares (CRLS) regression algorithm, which uses the recorded EOG channels (Fp1 and Fp2) as a reference to adaptively estimate and subtract ocular noise from the EEG data. Next, electrodes with excessive noise were identified using the ~\textit{clean\_rawdata} function in EEGLAB, whereby an electrode was considered abnormal if it flatlined for $>$5 s, had values exceeding 4 standard deviations of the total channel population, or failed to correlate at r $>$ .85 with a reconstruction of it based on other channels. Thresholds were selected based on standard default recommendations for ~\textit{clean\_rawdata}.

Electrode rejection rates were 6.5 ± 2.6 electrodes (mean ± SD). For alpha lateralization analyses, noisy electrodes were interpolated to ensure a consistent electrode array across all participants, enabling the calculation of alpha lateralization indices using standardized electrode clusters (Section~\ref{subsec:Alpha_Band_Lateralization}). However, these noisy electrodes were excluded from all IEM analysis to prevent the inclusion of potentially inaccurate reconstructed electrode data in the model (Section~\ref{subsec:Alpha_IEMs}). 
The data were then epoched from 0.5 s prior to stimulus onset to 2 s post-stimulus offset and the mean pre-stimulus baseline (-0.2 - 0 s) was subtracted. Trials with unsuccessful saber strikes were removed. A threshold-based artifact rejection function was then used to eliminate epochs in the remaining electrodes that exhibited amplitudes surpassing ± 150 µV (Trial rejection rates per condition: C1 = 5.6 ± 7.5 \%, C2 = 6.9 ± 9.3 \%, C3 = 4.9 ± 6.6 \%, C4 = 6.4 ± 8.8 \%). 

For the lateralization and IEM analyses, we first submitted the cleaned, epoched data to a third-order Butterworth filter to isolate the alpha (8-12 Hz) frequency band. A Hilbert transformation was then applied to extract instantaneous amplitude and phase values. Data were then sorted by location bin. To mitigate bias due to potential trial number discrepancies caused by artifact rejection, trial numbers were equalized across bins. The bin with the fewest trials was identified, and an (n-1) sized subsample was randomly selected from each bin. This ensured every bin contributed a random subsample, including the one with the least trials. Total alpha power (non-phase-locked oscillations) was then computed by squaring the absolute values from the Hilbert transform.

\subsection{Event-Related Potentials}
\label{subsec:ERPs}

We used the ERP technique to analyze time-locked neural responses to targets in the Static conditions. This was possible only for the Static conditions because those tasks featured abruptly onsetting lateralized targets. Specifically, we examined the visual N1 and N2pc components - negative deflections occurring at parieto-occipital scalp locations between ~150-300 ms after the  presentation of a visual stimulus that provide precise time=locked signatures of spatial attention ~\cite{eimer_n2pc_1996,johannes_luminance_1995}. We isolated these components by averaging brain responses across trials at electrodes both contralateral (i.e., opposite side) and ipsilateral (i.e., same side) to the location of the target sphere. Consistent with standard guidelines for measuring lateralized visual attention components ~\cite{luck2014introduction} , we analyzed both components at lateral occipito-parietal electrodes PO7/PO8 and P7/P8, where they are typically maximal.


\subsection{Lateralization}
\label{subsec:Alpha_Band_Lateralization}

We examined alpha-band lateralization as this provides a moment-by-moment measure of spatially selective attention that, unlike the ERP analyses described in the previous section, is not strictly time-locked to abrupt stimulus onsets. This allows us to measure attention in the static conditions but, also critically in the dynamic conditions, where we want to attend to objects that start at fixation and gradually move into the periphery. 

Alpha oscillations are extensively involved in voluntary attentional allocation ~\cite{foxe_role_2011}, such that the distribution of alpha power across parieto-occipital scalp regions changes when attention is focused on stimuli located on either side of the visual field \cite{sauseng_shift_2005, thut_alpha-band_2006}. To examine the role of alpha oscillations in our data we computed a lateralization index ~\cite{thut_alpha-band_2006,harris_distinct_2017}. We first averaged alpha power across parieto-occipital and occipital electrodes where alpha lateralization is known to be most evident [~\cite{thut_alpha-band_2006,sauseng_shift_2005}; left ROI: PO3, PO7, O1; right ROI: PO4, PO8, O2]. Then, we computed the degree of lateralization over the trial duration using the following equation:

$$Lateralization Index = \frac{Ipsilateral Power - Contralateral Power}{Ipsilateral Power + Contralateral Power}$$

\subsection{Inverted-Encoding Modeling}
\label{subsec:Alpha_IEMs}

\begin{figure}[t] 
  \centering 
  \includegraphics[width=1\columnwidth]{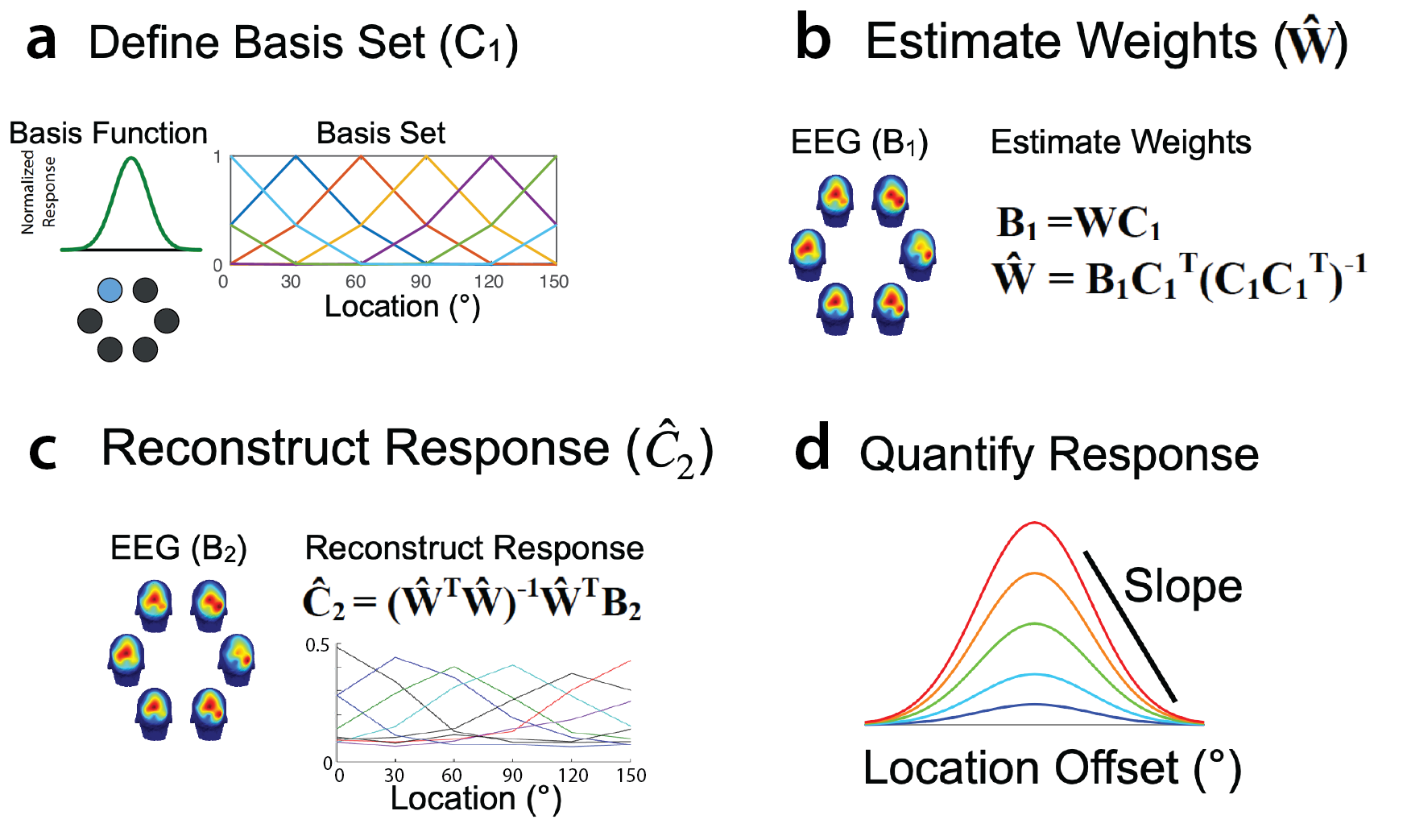}
  \caption{Inverted Encoding Modeling (IEM) Sequence. The IEM assumes brain responses can be modeled as a set of hypothetical information channels~\cite{sprague_assaying_2018}. The first step is to estimate to what extent a linear combination of predetermined channel responses (the basis set, (a)) captures the underlying structure of the observed EEG data. This process generates regression weights for each electrode and channel, which essentially represent the contribution of each electrode to each location-specific channel (b). Next, these weights are used to estimate channel responses from the observed EEG data (c). Spatially selective responses are then quantified by circularly shifting the CRFs to a common center (0 degrees), folding them at the center and computing the linear slope (d).}
  \vspace{-2ex} 
  \label{fig:Fig03_IEM_Methods}
\end{figure}

Here, we used IEM to estimate spatially selective neural responses based on topographical patterns of alpha activity across the scalp ~\cite{samaha_decoding_2016,foster_topography_2016,maclean_dual_2019, garrett_tracking_2021}.
An IEM attempts to determine whether the input features are carrying information with specific characteristics for each of the stimulus conditions and, in the process, the method can be used to reconstruct a specific feature that the brain is representing at a moment in time. 
This provides a quantitative estimation of whether the information conforms to an specific expected pattern. A visualization of the IEM pipeline is shown in (Figure \ref{fig:Fig03_IEM_Methods}) and more detail about each step is described below. 

We prepared the data for the IEM by first organizing the spectral data into location bins, ensuring the number of trials per bin was equal. To enhance the signal-to-noise ratio and reduce computational demands, each bin was then randomly split into three sets of trials, which were averaged to create 18 total inputs for the IEM (6 locations x 3 averaged trials). This trial selection process was repeated 10 times, with a new, random set of trials in each iteration, to ensure the model's results were not influenced by any specific set of trials. This robust approach improved the reliability of our model's outcomes.

An IEM was run for every time point within the 2.5 s trial epoch, resulting in 625 (250 Hz x 2.5 s) separate models per condition and participant. For model validation, we used an k-fold cross-validation approach, where \textit{k} = 4. We divided the averaged trials into four folds, with each fold containing one averaged trial per location bin. The model was trained on three of these folds and then tested on the single remaining "left-out" fold. This process was repeated four times, ensuring that each fold was used once as the test set.

We ran two separate models: one for the static conditions and one for the dynamic conditions. Each model was trained using an equal number of trials from both conditions within its pair. This "fixed" encoding model approach helps to minimize potential bias from variations in signal-to-noise ratio (SNR) between conditions~\cite{liu_inverted_2017,sprague_assaying_2018}. For each participant and condition, we defined the following variables: \textit{m} represents the number of EEG electrodes; $n_\textit{1}$ and $n_\textit{2}$ represent the number of trials in the training and testing sets, respectively. We then define a basis set. Let \textit{j} be the number of hypothetical response channels $(C_\textit{1}, j \times n_\textit{1})$, comprised of half-sinusoidal channels raised to the seventh power. This basis set was comprised of six channels representing equally-spaced locations (i.e., $j$=6). The training and test sets are represented by $B_\textit{1}(m \times n\textit{1})$ and $B_\textit{2}(m \times n\textit{2})$, respectively. A general linear model (GLM; standard implementation) was then used to estimate a weight matrix $(W, \textit{m} \times \textit{j})$ using the basis set $(C_\textit{1})$. Specifically:

$$B_1 = WC_1$$
 
The ordinary least-squared estimate of W can be computed as follows:
 
$$\hat{W} = B_1C_1^T(C_1C_1^T)^{-1}$$
 
Using the estimated weight matrix ($\hat{w}$) and the test data ($B_\textit{2}$), the channel responses $(C_\textit{2}, j \times n_\textit{2})$ can be estimated by the following:
 
$$\hat{C}_2 = (\hat{W}^T \hat{W})^{-1} \hat{W}B_2$$

After calculating the estimated channel responses $\hat{C}_2$ for each location bin, we circularly shifted the channel response function (CRF) for each averaged trial to align with a stimulus-centered reference frame. Centered CRFs were then averaged across trials for each participant and condition. This entire modeling process was repeated for every time point in the trial. To ensure the model's output was not biased by any particular trial selection, we ran the model ten times, each with a different random selection of trials. A final centered CRF was then generated by averaging results across the ten iterations.

The IEM sequence was also repeated with randomly permuted location bin labels. We ran this null model ten times, each time using a different set of randomly shuffled bin labels, following the same trial randomization protocol. This entire process was repeated five times using a new set of shuffled labels each time. The resulting CRFs were averaged across all iterations and permutations to generate "permuted" CRFs. This randomization is expected to remove any spatial selectivity, resulting in flat channel response profiles. Permuted CRFs then served as a baseline for hypothesis testing, allowing comparison of "real" CRFs against a null model with no spatial selectivity.

To quantify spatial selectivity, estimated CRFs were then folded around 0° offset. This transformed the responses from [-120°, -60°, 0°, 60°, 120°, 180°] into [0°, 60°, 120°, 180°] by averaging responses at corresponding offsets (note that ±60° and ±120° were averaged, but 0° and 180° were not). The slope was then calculated using a linear regression of the resulting alpha power across the offset values. Larger slope values indicate greater spatial selectivity.

\subsection{Hypothesis Testing} 
We used a non-parametric permutation-based resampling technique to empirically approximate null distributions for our statistical tests (e.g., ~\cite{foster_topography_2016, bullock_habituation_2023, bullock_electrophysiological_2025,bullock_effects_2021}). For pairwise analyses, we shuffled condition labels within subjects and ran 1,000 iterations to generate null distributions of t-statistics. For multivariate techniques, such as IEM, the model was trained and tested for 1,000 iterations, with condition labels shuffled in each iteration to create a matrix of permuted null channel response functions. We then determined statistical significance by calculating the probability ($p_{null}$) of obtaining our observed \textit{t}-statistics from these null distributions. The standard ~\textit{F} and ~\textit{t}-statistics are reported alongside the critical \textit{p}-value (labeled $p_{null}$). Where appropriate, effect sizes are reported as Cohen’s \textit{d}. We acknowledge that performing repeated comparisons across the time-course increases the risk of Type I errors. However, our conclusions are not based on isolated significant timepoints, but rather on sustained temporal clusters of significance. The consistency of these effects across contiguous windows mitigates the likelihood of spurious results arising from multiple comparisons.

\section{RESULTS AND DISCUSSION} 
We present an integrated results and discussion section to create a more coherent narrative. This approach allows us to immediately explain the significance of each key result alongside the data, directly connecting our findings to the existing literature.  We then synthesize the results in a General Discussion section and discuss their broader implications. 

The static and dynamic tasks differed significantly in nature, timing, and instructions. The static tasks required participants to respond as quickly and accurately to each target as possible, whereas the dynamic tasks required waiting until a target came within reach. Due to these fundamental differences in structure and timing, we analyze the static and dynamic tasks as separate pairs.

\subsection{Behavior} 
Accuracy was near ceiling in both Static conditions, hence meaningful statistical comparison of accuracy was not possible due to lack of variance. Responses were faster in the Static-Single compared to Static-Multiple condition $[t(31)=9.34, p_{null}<.001]$ (Figure \ref{fig:Fig04_Behavior}a). Accuracy was not statistically different across Dynamic-Single and Dynamic-multiple conditions $[t(31)=1.86, p_{null}>.05]$, but responses were slightly faster in the Dynamic-Multiple condition, albeit by $<$1ms $[t(31)=3.46, p_{null}<.01]$ (Figure \ref{fig:Fig04_Behavior}b). 


\subsection{Evoked Brain Responses to Static Targets}
\begin{figure} 
  \centering \includegraphics[width = 1\columnwidth]{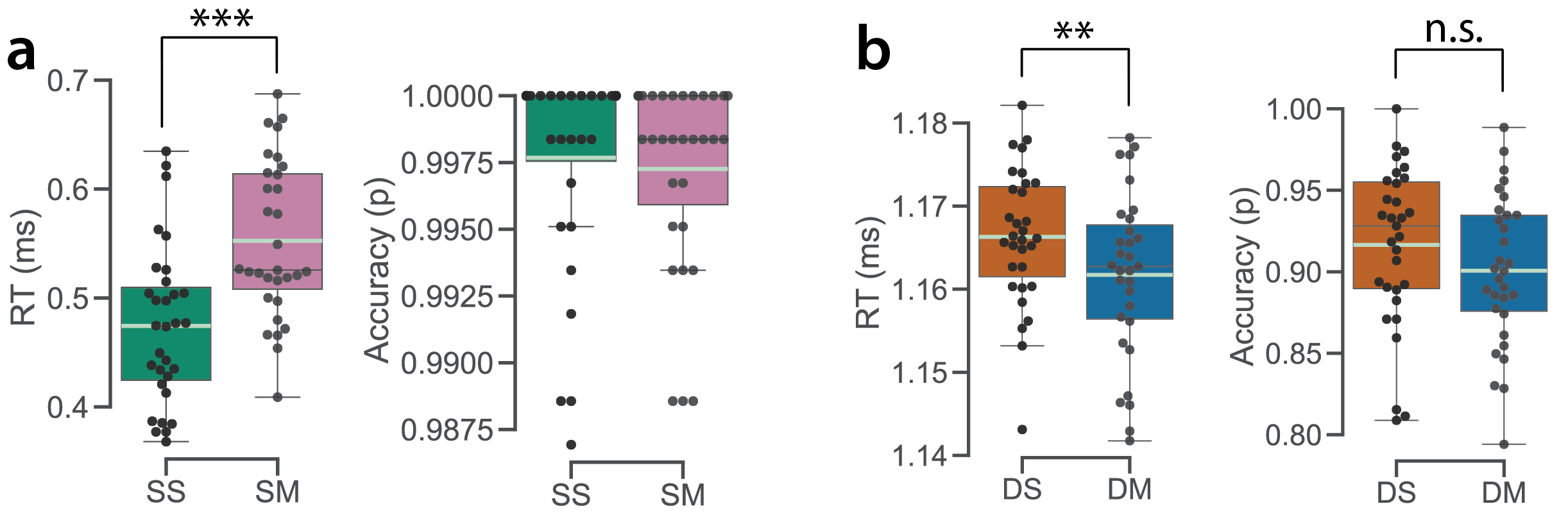}
  \caption{Behavior. Response times and accuracy for (a) Static-Single (SS) and Static-Multiple (SM) conditions, and (b) Dynamic-Single (DS) and Dynamic-Multiple (DM) conditions. Gray and white lines in boxplots represent group medians and means, respectively. **$p_{null} < .01$, ***$p_{null} < .001$}
  \vspace{-2ex} 
  \label{fig:Fig04_Behavior}
\end{figure}

\begin{figure} 
  \centering \includegraphics[width = \columnwidth]{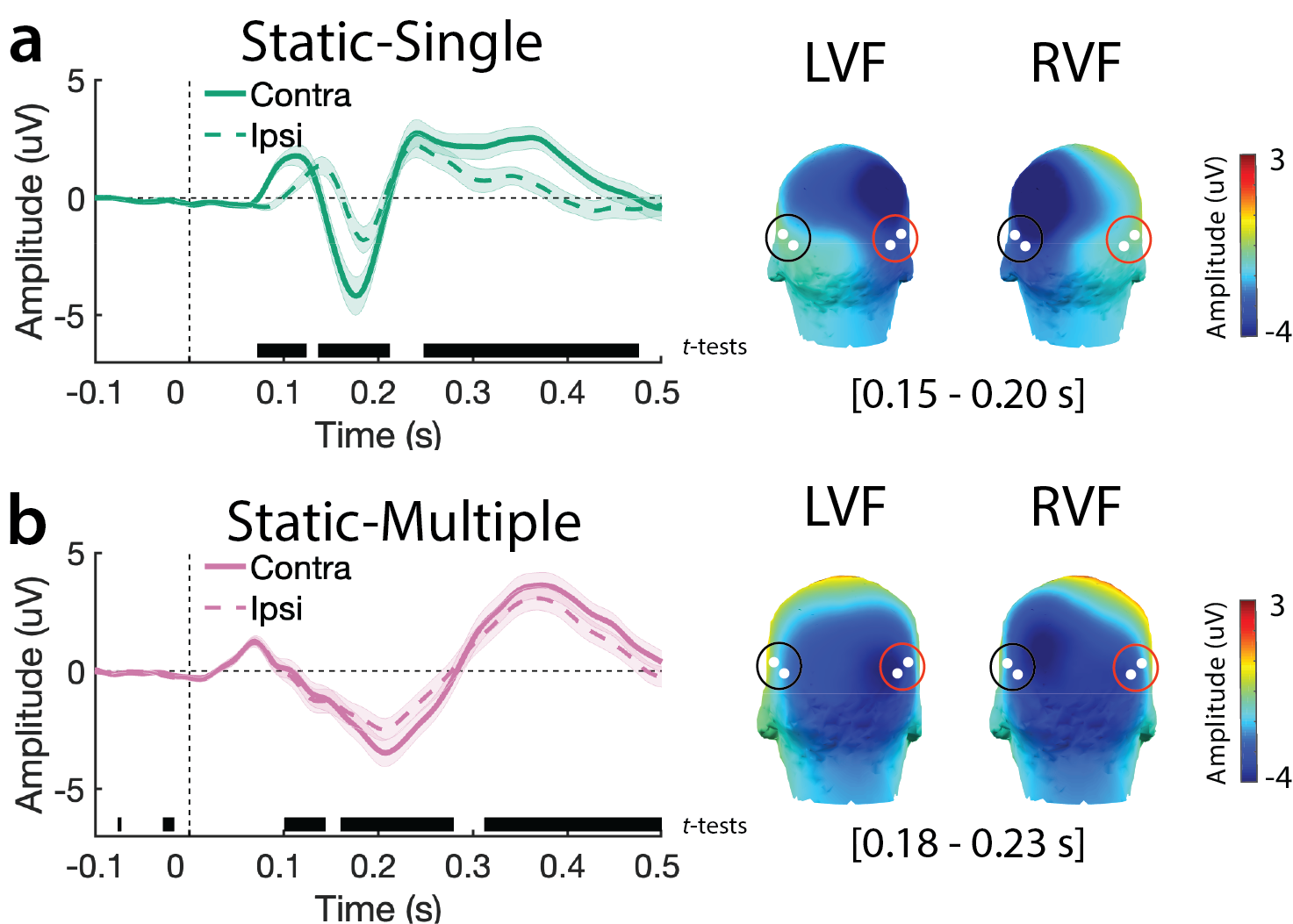}
  \caption{Event-Related Potentials. ERP Waveforms depict averaged contralateral and ipsilateral brain responses to target onsets in the Static-Single (a) and Static-Multiple (b) conditions. Horizontal black bars at the base of each plot mark timepoints where the lateralized responses are statistically different. Topographic maps to the right of the ERP plots show mean activation across posterior scalp regions for the large negative-going deflection averaged across the indicated timepoints. Contralateral and ispilateral electrodes used to compute waveforms are marked with red and black circles, respectively.}   
    \vspace{-2ex} 
  \label{fig:Fig05_ERPs}
\end{figure}

We analyzed ERPs time-locked to the onset of static targets to confirm that our 3D VR stimuli produced brain responses consistent with those observed in classic 2D studies of spatial attention. Inspection of contralateral and ipsilateral responses across posterior electrodes to targets presented either in isolation or with distractors revealed substantial differences between conditions (Figure \ref{fig:Fig05_ERPs}). The initial positive deflection around 100 ms post-stimulus (P1 ERP component) in both conditions is thought to reflect early sensory processing and attentional allocation in the brain ~\cite{hillyard_sensory_1998}. In the Static-Single condition, this was a highly lateralized response because the target appeared in isolation, leading to a rapid orientation of attention. In contrast, the Static-Multiple condition, which required a sequential search, showed a non-lateralized P1 because the onset of all six stimuli drove the response. These patterns are broadly consistent with previous work ~\cite{Clark1994}. Both conditions then show large subsequent negative-going lateralized deflections (i.e., larger negative-going amplitude in contralateral than ipsilateral). These deflections are broadly consistent with the posterior lateralized N1 and N2pc components, which are thought to reflect sensory gain control processes and the enhancement of the target signal from distracting information ~\cite{eimer_n2pc_1996,luck_electrophysiological_1994,wascher_visuospatial_2009}. The classic N2pc typically involves separation of the contralateral and ipsilateral waveforms at the second negative-going component in the waveform. In our data, the large negative deflection is the first and only negative component in the Static-Single condition, which is consistent with the visual N1; whereas in the Static-Multiple condition it is the third component. The difference in timing and also order likely reflect additional stages of attentional processing required for selecting a target amongst distractors (volitional, sequential search) versus selecting an isolated target (reflexive, no search). Changes in the timing and order of lateralized negative-going brain responses to spatial stimuli have also been observed in previous work ~\cite{eimer_n2pc_1996, verleger_time-course_2012}. Furthermore, the notion that search for a target amongst distractors requires additional stages of attentional processing is consistent with slower RTs in the Static-Multiple condition (Figure \ref{fig:Fig04_Behavior}a). 

\subsection{Lateralized Alpha Tracks Target Locations }
We examined lateralized responses in the alpha band to determine if these could be used to track the moment-by-moment allocation of spatial attention to both static and dynamic target objects presented in either left or right visual field locations. 

\begin{figure*}[t] 
  \centering 
  \includegraphics[width= .95\textwidth]{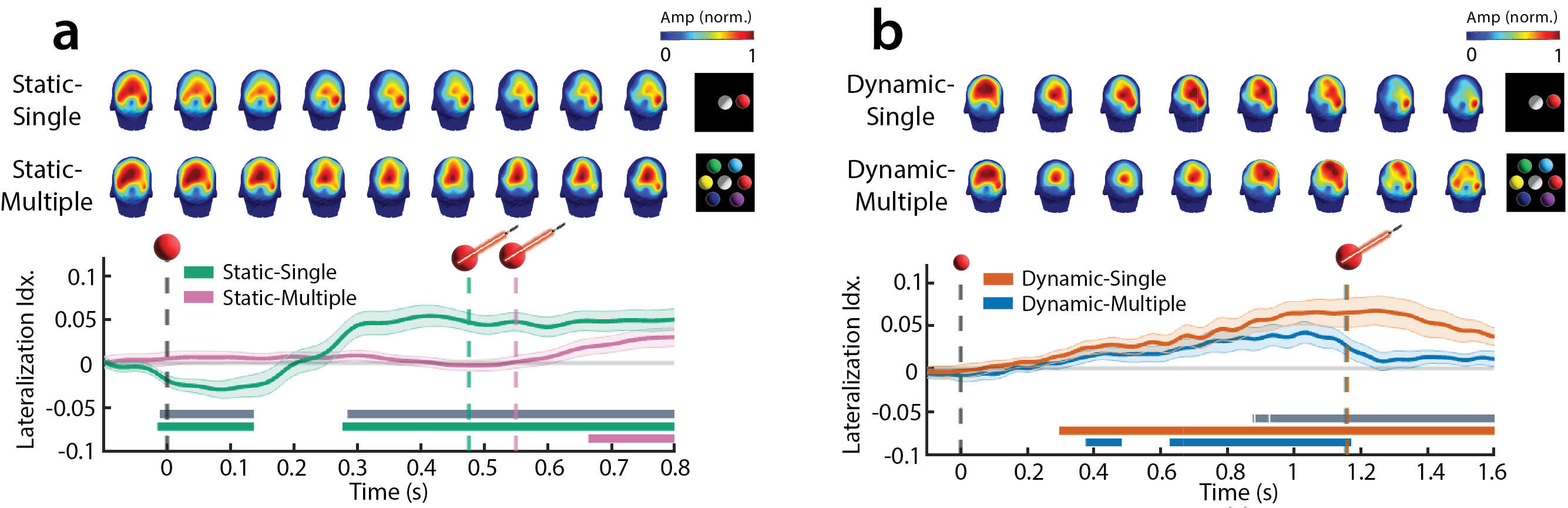}
  \caption{Alpha Lateralization. Plots show lateralization for (a) Static and (b) Dynamic conditions. Topographic maps show the distribution of alpha across posterior scalp sites related to one example target location (red target, 90\degree). At the base of each plot, condition-specific color-coded bars indicate timepoints where the lateralization index was significantly different from zero, while grey bars mark time points where lateralization differed significantly between conditions. Additionally, condition-specific color-coded vertical dashed lines indicate the mean saber strike reaction times for each condition.}
    \vspace{-3ex} 
  \label{fig:Fig06_Lateralization}
\end{figure*}

In the Static-Single condition, alpha power increased contralateral to the cue at target onset, then shifted ipsilateral to the cue ($\sim$0.3 s) and remained lateralized well beyond the mean RT (Figure \ref{fig:Fig06_Lateralization}a). The observation of alpha lateralization appearing slightly before stimulus onset is interpreted as an artifact of temporal smearing caused by the Butterworth filter. The later ipsilateral shift was expected and indicates allocation of attention to the attended location, consistent with previous work (e.g., ~\cite{thut_alpha-band_2006,sauseng_shift_2005}). The earlier contralateral activation was unexpected and to our knowledge has not been observed previously in screen-based spatial tasks. In the Static-Multiple condition alpha did not lateralize until later in the trial ($\sim$0.65 s), indicating lateralization is delayed if search is delayed. 



In the Dynamic-Single condition, alpha shifted ipsilateral from the target from $\sim$0.3 s and the lateralization index continued to increase until around the time of the response, where it plateaued before starting to decline (Figure \ref{fig:Fig06_Lateralization}b).  The monotonic increase suggests that alpha tracks the location of the target as it approaches from the center of the visual field and moves towards the periphery. In the Dynamic-Multiple condition alpha lateralized slightly later ($\sim$0.4 s to  0.5 s), before dropping, then building up again until the response ($\sim$.625 to 1.175). The delayed lateralization likely reflects the challenge of initially identifying and tracking the target amongst non-targets. The earlier plateau and decline immediately prior to response may reflect attentional capture by the looming non-targets, drawing attention from the target and causing it to be spread elsewhere.  

\subsection{Reconstructing Attended Target Locations} 

\begin{figure*}[t] 
  \centering 
  \includegraphics[width= .96\textwidth]{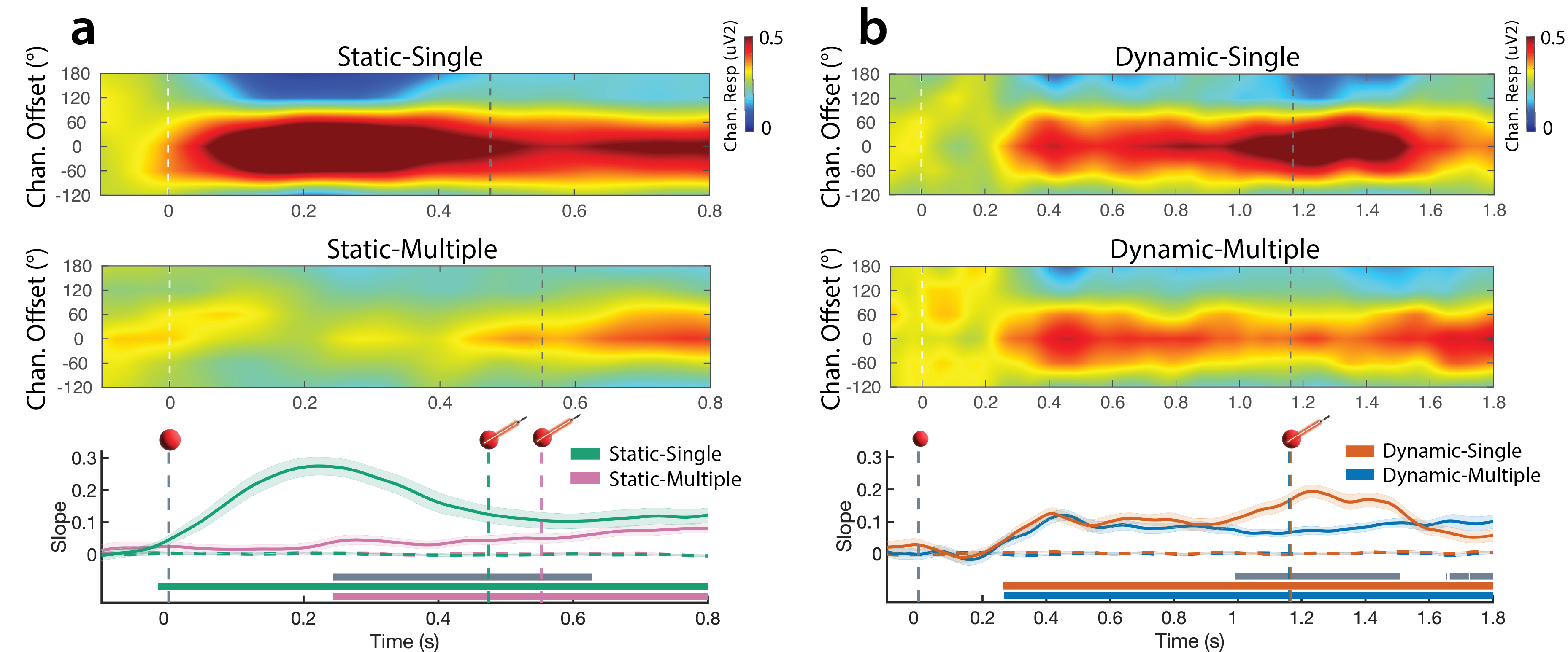}
  \caption{Reconstructing Object Locations from Alpha-Band Power with Inverted Encoding Modeling. CRF heatmaps and slope plots are shown for Static (a) and Dynamic (b) conditions. Heatmaps depict spatially selective responses for targets reconstructed from topographic patterns of alpha, with darker red along the preferred location (0 degs) indicating stronger spatial selectivity. Line plots depict linear slopes of folded CRFs for each condition (solid lines) and their permuted counterparts (dashed lines). Higher slopes indicate more selective responses. At the base of each plot, condition-specific color-coded bars mark timepoints where there is a robust location-selective responses to the target (i.e., CRF slopes are significantly different to their permuted counterparts). Gray bars mark timepoints where CRF slopes differed significantly between conditions. Additionally, condition-specific color-coded vertical dashed lines indicate the mean saber strike reaction times for each condition.}
  \vspace{-3ex} 
  \label{fig:Fig08_IEM}
\end{figure*}

We examined location-selective responses to determine the extent to which the target location was represented in patterns of alpha activity during each time point in the trial. CRF heatmaps and slope plots allow us to quantify and compare the strength of this location-specific coding throughout the trial (Figure \ref{fig:Fig08_IEM}).

In the Static-Single condition, robust location-selective reconstructions were observed at target onset and endured well beyond the mean saber strike RT. As with our previous lateralization analysis, the observation of a location-selective reconstruction appearing slightly before stimulus onset is interpreted as an artifact of temporal smearing caused by the Butterworth filter. The CRF slopes peaked between $\sim$0.2 and 0.3 s before declining and leveling off (Figure \ref{fig:Fig08_IEM}a). In contrast, in the Static-Multiple condition, location-selective responses emerged much more slowly, with robust reconstructions emerging at $\sim$0.25 s and persisting until after the saber strike. Together, these data indicate that it takes participants longer to orient attention to targets surrounded by distractors compared to isolated targets. This is consistent with the previously presented behavioral and EEG data.  


In the Dynamic-Single condition, location-selective information was present from $\sim$0.25 s post-stimulus and was sustained through the end of the trial (Figure \ref{fig:Fig08_IEM}b). The CRF slopes initially peaked around 0.45 s and then peaked again just prior to target interception, before dropping off. In the Dynamic-Multiple condition, location-selective information also emerged from $\sim$0.25 s, peaked just slightly after the Dynamic-Single condition and then sustained through the end of the trial. Notably, CRF slopes diverged between the two conditions around the time of target interception ($\sim$1 to 1.5 s), with a clear peak in the Dynamic-Single relative to the Dynamic-Multiple condition, indicating stronger focus or orientation of attention to a lone target. This divergence may reflect diffusion of attention away from the target when it is surrounded by distractors around the time of interception, despite the target already having been identified earlier in the trial. This may reflect participants being unable to disengage attention from the looming distractor stimuli, suggesting these stimuli capture attention despite their task irrelevance. This is consistent with previous behavioral work showing capture by looming stimuli~\cite{lin_objects_2008, franconeri_moving_2003}.




\section{GENERAL DISCUSSION}

The goal of this paper was to assess the feasibility of the SABER framework for studying the neural mechanisms of object tracking in immersive VR. Participants used virtual sabers to strike target spheres, which appeared either abruptly or loomed from a distance, both alone and with distractors. Participants performed well at the tasks, striking targets with high accuracy. We then demonstrated that methods previously used to study spatial attention to static objects on 2D screens could be applied to study both static and moving 3D objects in immersive VR. Our results validate this approach by demonstrating a consistency in neural mechanisms with previous 2D studies while also providing novel insight into spatial attention to dynamic 3D objects. This paper thus provides a crucial foundation for future research, enabling new hypotheses about how the brain supports attention to dynamic objects to be tested.

Traditional univariate EEG methods confirmed that our static and dynamic 3D VR stimuli produced reliable brain responses consistent with the existing literature. The ERP analyses served as an important reality check on timing, showing that expected sensory-evoked responses unfolded on a plausible timeline following the abrupt onset of targets. This finding gives us confidence in the temporal precision of our integrated VR-EEG hardware, despite the temporal smearing inherent in oscillatory analyses. Furthermore, lateralization analyses confirmed that alpha power shifted across the posterior scalp as participants attended to targets, a pattern consistent with previous work. A key novel result from this analysis was demonstrating that, in the dynamic conditions, the extent of alpha lateralization tracks the target's eccentricity from fixation. Additionally, the unexpected rise in contralateral alpha power following target onset in the Static-Single condition represents a novel finding that has not been documented in prior literature. The exact cause of this effect is unclear, but we reason that this is a genuine physiological effect rather than a methodological artifact, as very similar processing pipelines yielded no such patterns in our previous screen-based experiments using comparable single-target paradigms ~\cite{bullock_eye_2023,maclean_dual_2019}. Replication and further examination of this unexpected effect will be important. Note that while we interpret lateralized alpha activity as a proxy for attentional allocation, the precise functional role of these oscillations remains a subject of debate. Although there is consensus that alpha tracks the spatiotemporal dynamics of covert orienting, current evidence cannot definitively distinguish between mechanisms of distractor suppression and target signal enhancement~\cite{foster_role_2019}. Alpha oscillations likely represent just one of several components of a multifaceted attention system, and while alpha does appear to support a causal role in control of spatial attention, the specific processing stages and temporal dynamics governing this effect remain to be fully characterized~\cite{peylo_cause_2021}.

Multivariate methods indicated that it was possible to track moment-by-moment spatial attention to 3D static and dynamic objects with greater spatial specificity. Unlike traditional univariate measures which often provide only coarse hemispheric contrasts (attend left vs. right), the IEM approach allowed us to recover precise spatial reconstructions of attended targets from the patterns of alpha oscillations recorded across the scalp. The IEM results are particularly encouraging for cognitive neuroscientists, as they show it is possible to reconstruct dynamic 3D target locations despite the increased visual complexity and potential noise inherent to VR environments. This capability is essential for studying the dynamics of spatial attention, paving the way for future research into how the brain continuously updates priority maps during naturalistic 'attentional smooth pursuit' of dynamic targets moving in 3D.


\subsection{Limitations and Future Directions}

While this study maintained high experimental rigor, we acknowledge several limitations that could be addressed in future versions of SABER. First, to isolate covert attention and minimize ocular artifacts known to disrupt alpha-band activity, we required participants to maintain central fixation~\cite{maclean_dual_2019,bullock_eye_2023}. We recognize that enforcing fixation limits ecological validity, as natural attention involves overt eye movements; however, this constraint was necessary to minimize noise in this proof-of-concept study and allowed for direct methodological comparison with foundational screen-based paradigms~\cite{eimer_n2pc_1996,samaha_decoding_2016, thut_alpha-band_2006}. Aspects of gaze control could be refined depending on future research goals. To ensure stricter control in future versions of SABER aimed at isolating covert attention, a self-paced, gaze-contingent approach will be implemented, where trials automatically terminate upon broken fixation, thereby ensuring participants strictly adhere to the central fixation requirement. Conversely, future adaptations of the task targeting naturalistic, free-viewing behavior will eliminate fixation requirements, thereby enabling participants to overtly attend to targets. Further work will also strictly control low-level visual properties by equating stimuli for luminance and contrast. Second, our sample was predominantly composed of young, healthy, female adults, which may limit the generalizability of our findings. Future iterations will prioritize recruiting a demographically representative sample with a balanced sex ratio and broader age range, if possible. Third, despite its immersive nature, VR differs from reality and remains a novel experience for many participants, which may impact behavior. The extended duration of the VR session ($\sim$ 1 hour) likely contributed to participant fatigue, while the standing posture required for the task may have taxed attentional resources and introduced additional variability into the neural data ~\cite{fakorede_neural_2025,zinchenko_withstand_2025}. While we are not yet capturing true real-world behavior, this VR-EEG approach does go a long way towards bridging the gap between traditional lab-based studies and naturalistic human functioning. Fourth, despite the generally clean EEG signals, the VR headset design may have compromised electrode contact in specific areas, notably under the overhead strap and face pad. Fifth, the HTC Vive Pro Eye has a physical minimum IPD of 60.7 mm. As our sample included participants with IPDs sightly below this threshold (mean = 59.9 mm), a subset of participants viewed stimuli with a slight optical misalignment. While this offset (typically $<$ .04 mm per eye) generally remains within the sweet spot of the Fresnel lenses, it may have introduced minor optical aberrations for participants with significantly smaller IPDs. While this is not ideal, since all comparisons were within-subjects, any minor optical discrepancies were consistent across conditions and did not introduce experimental bias. We also obtained verbal confirmation from participants that the stimuli were not blurry each time the headset was repositioned.

Despite limitations, this paper provides proof-of-concept for a new framework for studying attention in dynamic 3D VR. While the primary focus was validating SABER for spatial attention research, this approach also opens the door to investigating the interplay between attention, working memory, and action goals. Recent hypotheses on working memory propose that a neural mechanism exists to track stored items across space and time~\cite{awh_working_2025}, and our approach provides evidence to support this. Furthermore, considering accumulating evidence that action goals~\cite{handy_graspable_2003,trentin_action_2024,trentin_visual_2023}, movement, and physiological state~\cite{garrett_systematic_2024,giesbrecht_physically_2025,bullock_multiple_2015} influence attention and working memory, it is critical to adopt more dynamic and embodied tasks to gain a comprehensive understanding of these cognitive functions.

\section{CONCLUSION} 

Tracking dynamic objects relies heavily on spatial attention and WM ~\cite{belledonne_adaptive_2025, drew_neural_2008, holcombe_attending_2023} yet contemporary theories of attention and working memory are primarily based on static objects presented at fixed locations  ~\cite{Corbetta2002, fiebelkorn_functional_2020, luck_progress_2021, shomstein_attention_2023, wolfe_guided_2021}. This paper introduces and validates SABER, a new framework that merges a dynamic~\textit{Beat Saber} inspired VR task with EEG recording to elucidate the neural mechanisms that underpin real-time tracking of dynamic objects in an immersive 3D environment. We show how EEG techniques typically applied to study spatial attention to 2D static stimuli can be successfully used to study dynamic stimuli moving in 3D. These findings validate the SABER framework and open up new avenues for research into how the brain supports attention to dynamic objects. This approach reveals how known neural mechanisms for static object processing extend and adapt to dynamic contexts, while importantly allowing us to challenge the notion that attentional mechanisms operate identically in static versus dynamic settings ~\cite{djebbara_turning_2025}. This approach is critical because the insights it can generate will not only advance our understanding of how attention operates in the physical world but will also have practical applications in the design of more intuitive interfaces, effective training simulations, and immersive experiences optimized for the human attention system. Ultimately, this work is relevant across a broad range of fields, including cognitive neuroscience, human factors, and extended reality (XR) and brain-computer interface (BCI) development.




\acknowledgments{
The authors wish to thank Elise Kormos, Paola Belmontes and Catherine Wen for their assistance with data collection.This research was sponsored by the U.S. Army Research Office and accomplished under contract W911NF-09-D-0001 for the Institute for Collaborative Biotechnologies. Additional support came from NSF award IIS-1911230 and ONR grant N00014-23-1-2118.}

\bibliographystyle{abbrv-doi}

\bibliography{references_NSF}

\end{document}